\def\PRD{{Phys. Rev.} D}
\documentclass[
    ,final            % use final for the camera ready runs
  ]
  {aipproc}

\layoutstyle{6x9}
\usepackage{bm}% bold math
\begin{document}

\title{Confinement in the \\lattice Landau Gauge QCD simulation}

\author{Sadataka Furui}{
  address={School of Science and Engineering, Teikyo University, Utsunomiya 320-8551,Japan}
}

\author{Hideo Nakajima}{
  address={Department of Information science, Utsunomiya University, Utsunomiya 320-8585,Japan }
}

%\author{<author3>}{
%  address={<common address for author2 and author3>}
%  ,altaddress={<author1 address>} % additional visiting address
%}

\begin{abstract}
The running coupling and the Kugo-Ojima parameter of the confinement criterion are measured for the quenched SU(3) $\beta=6.4, 6.45$, $56^4$ lattice and the unquenched $\beta=5.2$, $20^3\times 48$ lattice of JLQCD, $\beta=2.1$, $\kappa=0.1357, 0.1382$, $24^3\times 48$ lattice of CP-PACS and $\beta_{imp}=6.76,a m_{u,d}=0.007, 6.83,a m_{u,d}=0.040$, $20^3\times 64$ lattice of MILC collaboration. 

The quenched SU(3) $56^4$ lattice data suggest presence of infrared fixed point of $\alpha_s(0)=2.5(5)$ and the approach of the ensemble of the 1st copy to the Gribov boundary. The running coupling of $q>2$GeV can be fitted by the perturbative QCD(pQCD) + $c_1/q^2$ correction. We find the Kugo-Ojima parameter $u(0)=-0.83(3)$. 

The rotational symmetry of the gluon propagator of the unquenched SU(3) is partially recovered, but its magnitude depends on whether the Wilson fermion or the Kogut-Susskind(KS) fermion are coupled to the gauge field.  When the sea quark mass is sufficiently light, both fermions suggest infrared fixed point $\alpha_s(0)\sim 2-2.5$.  The Kugo-Ojima parameter of unquenched configurations with light fermion masses is consistent with $u(0)=-1.0$.  
\end{abstract}

\maketitle

%%%%%%%%%%%%%%%%%%%%%%%%%%%%%%%%%%%%%%%%%%%%
%% MAINMATTER
%%%%%%%%%%%%%%%%%%%%%%%%%%%%%%%%%%%%%%%%%%%%

\section{Introduction}

 In Landau gauge $\widetilde{MOM}$ schme, we measure the QCD running coupling in terms of gluon dressing functiuon $Z_A(q^2)$ and ghost dressing function $G(q^2)$, $\displaystyle \alpha_s(q)=\frac{g^2}{4\pi}\frac{G(\tilde q^2/a^2)^2 Z_A(q^2)}{\tilde Z_1^2}$. ($q=\sqrt{\tilde q^2+\tilde q^4(\delta/3)}/a$, $\delta=0$ for ordinary action and $\delta=1$ for the improved action\cite{cppacs,milc}.) It is a renormalization group invariant quatity, but in the finite lattice, the vertex renormalization factor $\tilde Z_1$ is not necessarily equal to 1 as in pQCD. We fix this value by the fit of the numerical result to the pQCD.

Colour confinement in infrared QCD is characterized by the Kugo and Ojima 
parameter $u(0)=-c.$ The parameter $c$ is related to the renormalization factor as $1-c=\frac{Z_1}{Z_3}=\frac{\tilde Z_1}{\tilde Z_3}.$
If the finiteness of $\tilde Z_1$ is proved, divergence of $\tilde Z_3$ is a
sufficient condition. If $Z_3$ vanishes in the infrared, $Z_1$ should have higher order 0.

\section{The ghost propagator and the gluon propagator }
The ghost propagator is the Fourier transform of an expectation value of the inverse Faddeev-Popov operator ${\cal  M}=-\partial D=-\partial^2(1-M)$
\begin{equation}
D_G^{ab}(x,y)=\langle {\rm tr} \langle \Lambda^a x|({\cal  M}[U])^{-1}|
\Lambda^b y\rangle \rangle
\end{equation}
where the outmost $\langle\rangle$ denotes average over samples $U$.

The ghost propagators $D_G(q^2)$ of quenched and unquenched SU(3) in $\widetilde{MOM}$ scheme can be fitted by the pQCD in $q>0.4$GeV region, and there is no $\beta$ dependence. It means that there are strong correlation between the string tension and the ghost propagator. In this presentation the gauge field of $\log$-$U$ type\cite{FN04} is adopted. The ghost propagators are 14\% larger when the $U$-linear definition is adoped.

The gluon propagator of $56^4$ lattice is finite at 0 momentum and $Z_3(0)$ is not compatible to 0 in the present lattice size.
The gluon propagator of the unquenched SU(3)\cite{jlqcd,cppacs,milc} is measured by adopting the cylinder cut. Extraction of infrared physical quantities becomes difficult due to lack of symmetry of the four coordinate axes.

We observe reflection positivity violation in the unquenched gluon propagator, and in some polarization components of sample-wise quenched gluon propagator.  The reflection positivity violation and the closeness of the Kugo-Ojima parameter to -1 are correlated.

\section{The QCD running coupling and the Kugo-Ojima parameter}
\begin{figure}[htb]
\begin{minipage}[b]{0.60\linewidth} 
%\begin{center}
\includegraphics{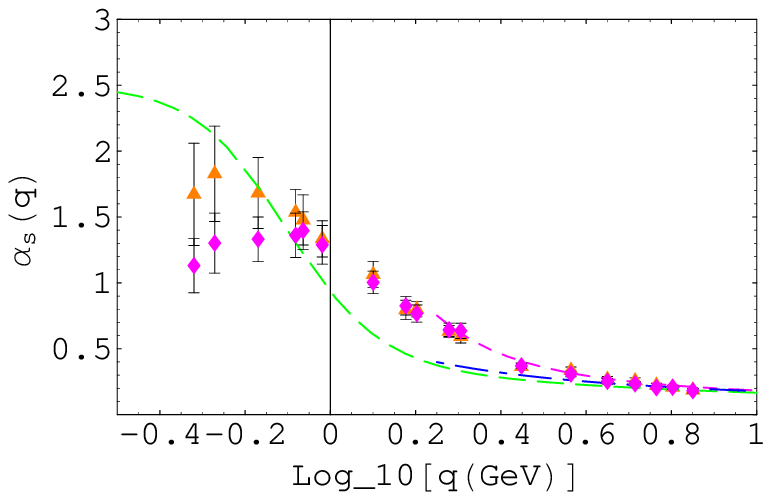}
%\end{center}
\end{minipage}
\hfill
\begin{minipage}[b]{0.60\linewidth}
%\begin{center}
\includegraphics{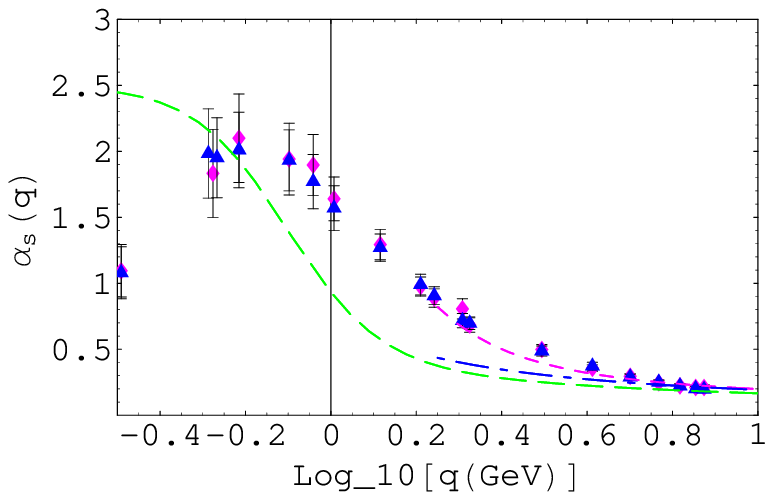}
%\end{center}
\caption{The running coupling $\alpha_s(q)$ of the CP-PACS of $K_{sea}=0.1357$ (diamonds) and that of 0.1382 (triangles). (25 samples) [left] and that 
of the MILC $\beta_{imp}=6.83, a m_{u,d}=0.040$(diamonds) and 6.76,$a m_{u,d}=0.007$(triangles). (50 samples)[right]. The DSE approach with $\alpha_0=2.5$(long dashed line), pQCD (dash-dotted line) and the Orsay fit of pQCD+$c_1/q^2$(dashed line) are also shown.}
\end{minipage}\label{alpunquench}
\end{figure}

The QCD running coupling  $\alpha_s(q)$ of the quenched SU(3) normalized at high momentum region by the 3-loop pQCD is close to the prediction of Dyson-Schwinger calculation of $\kappa=0.5$. The dependence of the running coupling on the sea quark mass of Wilson fermion (JLQCD, CP-PACS) and that of KS fermion(MILC) are qualitatively different. Relatively heavy Wilson fermion (JLQCD) does not show deviation from pQCD in $q>3$GeV region, but both fermions suggest infrared fixed point $\alpha_s(0)\sim 2-2.5$, when the sea quark mass is light. 
The left-most momentum point in the Figure \ref{alpunquench} is to be excluded in the cone-cut due to the finite size effect. 

The Kugo-Ojima parameters are summarized in Table 1. The parameter $c$ becomes larger as the lattice size becomes large. It is about 0.83(3) in $56^4$, while in the unquenched simulation it is consistent to 1, when the fermion mass is light. In the case of $U$-linear definition of the gauge field,  $c$ is about 10\% smaller than that of the $\log$-$U$ definition.

\begin{table}[htb]
\caption{The Kugo-Ojima parameter along the spacial directions $c_x$ and that along the time axis $c_t$ and the average $c$, trace divided by the dimension $e/d$, horizon function deviation $h$ of quenched Wilson action (The first two rows),  unquenched Wilson action (The second two rows ), unquenched Wilson improved action (The third two rows), and unquenched KS improved action (The last two rows).  }\label{hmctab}
%\begin{center}
\begin{tabular}{c|c|c|c|c|c}
$K_{sea}$ or $\beta$ & $c_x$     & $c_t$    &$c$ &  $e/d$        &    $h$     \\
\hline
$\beta=$6.4 & 0.827(27)&  0.827(27)&  0.827(27)& 0.954(1) & -0.12(3)\\
$\beta=$6.45 & 0.809(81)& 0.809(81)&  0.809(81)& 0.954(1) & -0.15(8)\\
\hline
$K_{sea}=$0.1340 & 0.887(87)  & 0.723(38) &0.846(106) & 0.930(1) & -0.084(106)  \\
$K_{sea}=$0.1355 & 1.005(217) & 0.670(47) &0.921(238) & 0.934(1) & -0.013(238)  \\
\hline
$K_{sea}=$0.1357 & 0.859(58)  & 0.763(36) &0.835(68) & 0.9388(1) & -0.104(58)  \\
$K_{sea}=$0.1382 & 0.887(87) & 0.723(38) &0.846(106) & 0.9409(1) & -0.051(87)  \\
\hline
$\beta_{imp}=$6.76 & 1.040(111)  & 0.741(28) &0.965(162) & 0.9325(1) & 0.032(162)  \\
$\beta_{imp}=$6.83 & 0.994(141)  & 0.748(32) &0.932(163) & 0.9339(1) & -0.002(163)  \\
\hline
\end{tabular}
%\end{center}
\end{table}
\section{Summary}
We observed that the running coupling of unquenched Landau gauge QCD in the infrared is enhanced when the sea quark mass is light and the $c_1/q^2$ correction which appeared in the quenched simulation appears also in the unquenched simulation. The KS fermion of MILC collaboration (Asqtad action) has smaller sea quark mass dependence than that of Wilson fermion of CP-PACS collaboration.

%%%%%%%%%%%%%%%%%%%%%%%%%%%%%%%%%%%%%%%%%%%%%%%%
%% BACKMATTER
%%%%%%%%%%%%%%%%%%%%%%%%%%%%%%%%%%%%%%%%%%%%%%%%

%\begin{theacknowledgments}
This work is supported by the KEK supercomputing project No.04-106. H.N.is supported by a JSPS Grant-in-Aid for Scientific Research on Priority Area No.13135210.
%\end{theacknowledgments}

%% If the bibliography is produced without BibTeX comment out the

\bibliographystyle{aipprocl} % if natbib is missing

%\bibliography{sample}

\begin{thebibliography}{99}
\bibitem{FN03} S. Furui and H.Nakajima, {\PRD}{\bf 69},{074505}{(2004)}, hep-lat/0305010 and references therein.
\bibitem{FN04} S. Furui and H.Nakajima, {\PRD}{\bf 70},{094504}{(2004)}, hep-lat/0403021 and references therein.
\bibitem{FN} S. Furui and H. Nakajima, in {\it Confinement IV}, Ed. W. Lucha et.al., World Scientific, Singapore, p.275(2002), hep-lat/0012017.
\bibitem{jlqcd}  S.Aoki et al., (JLQCD collaboration),{\PRD}{\bf 65},{094507}{(2002)}.%{\PRD}{\bf 68},{054502}{(2003)}.
\bibitem{cppacs} A. AliKhan et al., (CP-PACS collaboration),{\PRD}{\bf 65},{054505}{(2002)}.
\bibitem{milc} C.W. Bernard et al., (MILC collaboration), {\PRD}{\bf 64},{054506}{(2001)}.

\end{thebibliography}

\IfFileExists{\jobname.bbl}{}
 {\typeout{}
  \typeout{******************************************}
  \typeout{** Please run "bibtex \jobname" to optain}
  \typeout{** the bibliography and then re-run LaTeX}
  \typeout{** twice to fix the references!}
  \typeout{******************************************}
  \typeout{}
 }

\end{document}